# Four small puzzles that Rosetta doesn't solve

Rhiju Das

Department of Biochemistry and Department of Physics, Stanford University, Stanford, California USA

Address correspondence to: Phone: (650) 723-5976. Fax: (650) 723-6783. E-mail: rhiju@stanford.edu.

## Abstract

A complete macromolecule modeling package must be able to solve the simplest structure prediction problems. Despite recent successes in high resolution structure modeling and design, the Rosetta software suite fares poorly on small protein and RNA puzzles, some as small as four residues. To illustrate these problems, this manuscript presents Rosetta results for four well-defined test cases: the 20-residue mini-protein Trp cage, an even smaller disulfide-stabilized conotoxin, the reactive loop of a serine protease inhibitor, and a UUCG RNA tetraloop. In contrast to previous Rosetta studies, several lines of evidence indicate that conformational sampling is not the major bottleneck in modeling these small systems. Instead, approximations and omissions in the Rosetta all-atom energy function currently preclude discriminating experimentally observed conformations from de novo models at atomic resolution. These molecular "puzzles" should serve as useful model systems for developers wishing to make foundational improvements to this powerful modeling suite.

## Introduction

The Rosetta modeling suite has enabled macromolecule structure prediction and molecular design with unprecedented accuracy and functionality (see, e.g., [1,2,3,4,5,6,7] and refs. therein). Nevertheless, Rosetta's algorithms continue to face limitations. Despite several near-atomic-resolution successes in favorable cases, rigorous tests in blind trials indicate that the general *de novo* prediction of even small proteins remains out of reach [8,9]. Similarly, the precise sculpting of polar environments or energetic balances necessary for designing efficient catalysts or allosteric switches remains challenging [4,5,10]. How can we achieve mastery over three-dimensional modeling and engineering?

Historically, Rosetta publications have focused on positive new developments, and rightly so, considering there are many. Nevertheless, this special collection of PLoS One manuscripts gives the Rosetta community an opportunity to present and dissect negative results. Individual labs have discovered many such results in recent years and found them useful for reflection; but these problems have not been disseminated widely and often come as surprises to new users.

In this short paper, I will take this opportunity to argue that the current Rosetta codebase has not yet achieved a critical step in its maturation into a general modeling tool: a confident and predictive understanding of the simplest and smallest macromolecule structure problems. If well-defined systems as small as four residues are not solvable by Rosetta, why have any confidence that 150-residue domains (or their massive complexes) are appropriate for *de novo* 3D modeling and eventually precision engineering? I will present four classes of simple but still perplexing small puzzles that provide tangible entry points – and, perhaps, shared model systems – for current and future enthusiasts hoping to establish a confident and rigorous foundation for Rosetta modeling.

Before describing the puzzles, some historical perspective is in order. It may seem self-evident that one should start modeling with the very smallest known sequences that take on well-defined 3D structures. For several reasons, such a completely reductionist approach has not been the mainstream strategy in the Rosetta community. First, in early days, most cases of naturally occurring ultra-small proteins (say, 30 residues or below) were considered irregular and perhaps ill-defined, lacking the clear α or β secondary structure and hydrophobic cores that are the hallmarks of larger protein domains. Thus, initial Rosetta studies from the mid-1990s focused on 50- to 100- residue protein sequences that formed regular, clearly well-defined structures [11,12], and these challenges were passed down from developer to developer as "in-house" benchmark sets [1,13]. Second, the global folds of smaller macromolecules may be less robust to inaccuracies in assumed energy function than larger systems, as there are fewer key packing interactions that specify small folds. Third, early modeling work involved coarse-grained molecule representations and energy functions; at this medium resolution (4-8 Å), most compact conformations of a small protein segment are indistinguishable from each other. Fourth, the focus on medium-resolution folds of larger systems has given useful insights, e.g., into assigning protein functions that have been subsequently validated by functional studies [14].

These historical reasons to avoid small systems are no longer as relevant. First, since the late 1990s, several very small protein systems have been discovered or engineered and then clearly demonstrated to attain precisely defined 3D structures (see, e.g., [15,16,17]). The expanding database of RNA and non-natural polymers (and the development of Rosetta code to model them [18,19]) further increases the number of such small, well-defined puzzles. In addition, *de novo* modeling of short irregular loops and small proteins regularly appear as sub-problems in blind prediction targets and in the design of catalytic sites, conformationally switchable segments, and structured peptides. Predicting the structural features of these small systems and sub-systems – and even modeling the fine energetic balance between alternative structures – is no longer something to be avoided but instead a key goal of many Rosetta developers.

Most importantly, many Rosetta developers are now striving for predictive power with Angstrom-level resolution [20], although fruitful insights and methods development at medium resolution continue (see, e.g., [21,22,23]). The new focus on high-resolution modeling is motivated by a shared belief: atomic accuracy appears necessary for a deep understanding of catalysis, drug design, and evolution. In a few, favorable cases, Angstrom-level modeling has been achieved [1,3,8], sometimes with limited experimental data [24,25,26]. Nevertheless, high-resolution success is rare rather than the norm, and this lack of general predictive power is typically blamed on the difficulty of complete conformational sampling at the 1-2 Å scale. At present, only small model systems offer the prospect of comprehensive sampling and are therefore the most stringent tests of Rosetta's assumptions and energy functions at high resolution. Due to their ubiquity, their functional importance, and their unique ability to test modeling at atomic resolution, small macromolecule modeling problems are an important and unsolved frontier for Rosetta modeling.

## Results

There are at least four kinds of problems at this "small puzzle" frontier, from mini-proteins without and with disulfides to protein loops and RNA motifs. As illustration, the following descriptions present a single model system of each kind. Each of the selected systems has been extensively characterized by numerous experimental structural and energetic methods. In particular, in each case, the free energy associated with the experimental conformation has been measured to be at least 3 kcal/mol more stable than the ensemble of unstructured states at room temperature. (The expected energy gap between the experimental conformation and *any individual* conformation of the competing ensemble is therefore expected to be (much) greater than 3 kcal/mol.) Note that the focus herein will be on recovering high-resolution features of the experimental models; thus an acceptable Cα RMSD should be 1 Å or lower, comparable to the differences between structures solved in different crystallographic space groups or with different binding partners. Further, the puzzle descriptions include discussion of side-chain conformations deemed experimentally stable and important for each molecule's fold and function.

I will summarize prior data and recent Rosetta modeling runs on these puzzles, using at least two different Rosetta conformational search strategies for each case. As per the guidelines for this Special Collection on RosettaCon 2010 science, an extensive methods section gives modeling details, including Rosetta command-lines and protocol capture, that will permit new developers to rapidly reproduce and assess this work.

*A. Mini-proteins*: *the Trp cage*.
"Mini-proteins" with sizes well under 30 residues are ideal systems for testing modeling tools and, indeed, are widely studied in the molecular dynamics (MD) community. The Trp cage is a particularly well-characterized mini-protein with a length of 20 residues, engineered by truncating and optimizing exendin-4 from gila monster saliva [16]. Several MD studies have recovered lowest-energy Trp cage conformations *de novo* that agree with the experimental NMR structure (see, e.g., [27,28,29,30]).

Rosetta *de novo* modeling, on the other hand, fails to solve this problem. While occasionally sampling a near-native conformation, Rosetta's fragment-assembly/all-atom-refinement protocol ("abrelax") favors a tight cluster of structures with a backbone within 2 Å Cα RMSD of the native conformation but with the molecule's central tryptophan side-chain in an incorrect rotamer (Figs. 1A & 2A). An independent sampling approach, Stepwise Assembly (see

Methods), which does not make use of fragments or a coarse-grained search phase, yields the same conformations as the lowest energy solution (Fig. 2A). Extensive optical thermodynamic characterization and NMR spectroscopy of Trp cage and several dozen variants [30] have revealed no evidence for this discrete alternative state.

*B. Structured peptides: the marine snail toxin GI.*
Peptides in snake, spider, and other venoms; mammalian and plant defensins; and extracellular signaling molecules form a second rich set of modeling puzzles. These molecules share folds and likely evolutionary lineage and have been optimized by evolution for high stability, precise folds, and, most importantly, small size. By making use of disulfide bonds, structured peptides can reach lengths smaller than those of (disulfide-free) mini-proteins. The α-conotoxin GI, isolated from the fish-hunting marine snail *Conus geographicus*, was one of the first of these tiny but potent sequences [31]. Despite containing only 13 residues, the peptide forms a highly stable fold with two disulfides, whose structure has been determined with diffraction data to 1.20 Å [32].

Even with a "cheat" – tight distance constraints that enforce the molecule's native disulfide pairing – Rosetta *de novo* modeling (abrelax) gives low energy models that disagree with the crystal structure in all non-helical regions (not shown). The Stepwise Assembly algorithm yields even lower energy models that are still highly discrepant (Figs. 1B & 2B; 2.8 Å Cα RMSD over 13 residues).

*C. Protein loops: the chymotrypsin inhibitor.*
The *de novo* building of loops excised from crystal structures offer another set of well-defined toy puzzles, with relevance to real-world practical problems such as comparative modeling or loop design. Some of these tests are surprisingly challenging. The chymotrypsin inhibitor from barley seeds displays a 10-residue protease-binding loop that appears visually irregular but is highly structured [33,34]– even in the absence of docking to the protease target site, this region is positioned with atomic precision by hydrogen bonds to two arginines extended from the molecule's main body.

Excision of this segment and subsequent Rosetta *de novo* loop modeling, leveraging kinematic closure (KIC) strategies [35], gives excellent convergence, with the 10 lowest energy models giving the same loop conformations (within 0.3 Å RMSD of each other). Unfortunately, this structure is an incorrect, collapsed loop with incorrectly positioned arginines (Fig. 1C). [Additional calculations with StepWise Assembly converge to similarly collapsed loops; these data are not shown here due to current differences in modeled degrees of freedom in KIC *versus* StepWise protocols.]

*D. RNA motifs*: *the most stable tetraloop, UUCG.*
A final, rich source of simple modeling puzzles comes from structured RNAs. These molecules fold back on themselves to form numerous double helices interconnected by so-called noncanonical motifs [36]. Many of these motifs are quite small, including the ubiquitous 4-residue tetraloops that cap off double helices to form hairpin folds [37]. Rosetta fares quite poorly in modeling an UUCG tetraloop hairpin *de novo*. (The sequence studied here differs in the stem sequence from a UUCG hairpin modeled previously [19] and shows lower energy non-native conformations.) The lowest energy models derived from Fragment Assembly of RNA with Full Atom Refinement (FARFAR) achieve none of the non-canonical base-pairing geometries or base-stacking interactions known from crystal structures [38]. Even lower energy solutions are uncovered by StepWise Assembly (Fig. 1D and Fig. 2D), still with poor all-atom RMSD (3.3 Å).

## Discussion

What is going wrong with Rosetta? Computational optimization procedures can give poor solutions due to either poor conformational search strategies or inaccurate optimization functions. Most recent Rosetta modeling papers have emphasized conformational search over hundreds of degrees of freedom as a critical, shared bottleneck in many problems [7,8,9,19,25]. However, conformational search is *not* the issue for the four small puzzles described above. In some cases, classic and novel search strategies produce nearly identical incorrect models (Fig. 2A); and, in fragment assembly approaches, independent modeling runs converge well (Figs. 2A, C, & D). The strongest evidence for efficient conformational sampling is that *de novo* models achieve lower energies than native models that have themselves undergone extensive optimization (involving equal amounts of high-performance computation; see Fig. 2, Table 1, and Methods).

If the Rosetta energy function gave an accurate portrait of *in vitro* folding, these non-native low-energy conformations would be expected to give energies at least 3 kcal/mol higher (instead of lower) than the optimized native models. Thus, at least for these four small puzzles, it is the poor discrimination of the Rosetta all-atom energy function that emerges as the critical problem. In previous studies, the difficulty of conformational sampling for larger problems as well the greater energy gaps attained in those problems likely masked flaws in the Rosetta energy function (see, however, [35]). Nevertheless, developers have recognized many shortcomings of the energy function (see also the Perspective in this issue), and inspection of puzzles A-D confirms a sizeable fraction of this list of problems. Rosetta's solvation model neglects many-body effects, nontrivial solvation structure oriented around polar groups, and "second shell" water effects; the model only weakly disfavors buried unsatisfied polar groups (Fig 1A) that are seldom observed in experimental structures [39]. Rosetta's hydrogen bond (H-bond) potential neglects the effects of charged atoms, (anti-)cooperativity within H-bond networks (Figs. 1B & C), and includes, by default, a dependence of H-bond strength on burial that is better modeled in the solvation term (Fig. 1C). Further, Rosetta ignores electrostatic interactions (besides H-bonds) and their screening, a likely important factor in stabilizing the UUCG hairpin (Fig. 1D) [40]. Finally, Rosetta does not currently permit rigorous estimation of a model's free energy, which has been suggested to be important for, e.g., the Trp cage structure [29].

How can these and additional issues be fixed? While there has been a long history of fine-tuning individual terms of the Rosetta energy function (much of it unpublished), these efforts have led to few substantial improvements in benchmarks or actual changes in the main codebase. One potential barrier is that the physics of solvation, H-bonds, and screened electrostatic interactions are strongly coupled to each other, and indeed are reflected in partly unified terms in most other molecular modeling force fields (see, e.g., [41,42]). Any fully consistent fix will require the guidance of – and perhaps a complete rewrite by – an expert developer with a comprehensive understanding of the Rosetta codebase and its hidden quirks. Further, some of the potential fixes, such as non-pair-wise energy function terms, may not be compatible with core features of the Rosetta package, such as the packer used for rapid side-chain optimization and design. Nevertheless, there is hope. The recent refactoring of Rosetta into object-oriented code greatly facilitates the creation and testing of novel energy functions [43]. The reorganization may also permit incorporation of libraries such as OpenMM [44] that implement independently developed energy functions containing physics (such as polarizable atoms[41]) missing in Rosetta.

As a final point, it is important to mention that the four puzzles described herein are not outliers but, rather, representative of inaccurate results that Rosetta finds for many small modeling problems. Rosetta similarly fails to achieve high resolution predictions for other mini-proteins, such as the pinWW domain and TrpZip [17]; for other well-characterized disulfide-stabilized peptides such as the sea anemone toxin BGK [45]; for other functional loops, including the highly stable trypsin-binding loop from the jumping cucumber *E. elaterium* [46]; and other functional RNA motifs, including the bulged-G motif [47] (unpubl. results,

R.D.). This paper has focused on four particular cases as specific illustrations for new Rosetta contributors, but future solutions to any of these four puzzles should also be validated on more extensive benchmarks including analogous mini-protein, loop, or RNA motif cases.

A complete macromolecule modeling package must necessarily be able to address the smallest structure prediction problems. It is both bad and good news that Rosetta fails at what appear to be the simplest high-resolution puzzles. The bad news is that Rosetta, perhaps the leading software package for 3D modeling and design, has fundamental limitations. But this is excellent news for current and future Rosetta developers; attaining confident solutions for the smallest modeling problems is an important goal whose pursuit involves interrogating the most basic rules of molecular self-assembly. The four puzzles presented herein offer well-defined entry points to developers who are interested in pursuing this fundamental path.

## Materials & Methods

### Command-lines for Rosetta

We summarize sequences and Rosetta command-lines for each of the four puzzles herein. Most of the calculations in the paper were carried out with Rosetta release 3.2; and all calculations will be implemented in the next Rosetta release. Remaining models were generated with the Rosetta codebase in the Das lab branch (revision number 40197), available to Rosetta developers (in the Rosetta Subversion repository at https://svn.rosettacommons.org/source/branches/das_lab/); this code will be gladly provided to other academic users upon request.

*(A) Trp cage*

The modeled sequence was for the most stable variant of the Trp cage: **DAYAQWLKDGGPSSGRPPPS**. De novo modeling was carried out with the following *ABRELAX* and *NATIVE_RELAX* command lines (see, e.g., [8]), using fragment files obtained from the Robetta fragment modeling server [48] with the "no homologs" option:

```
AbinitioRelax.<exe> —database <path to rosetta_database>  -fasta
2jof.fasta -native 2jof.pdb -frag3  aat000_03_05.200_v1_3.txt  -
frag9 aat000_09_05.200_v1_3.txt -out:file:silent 2jof_abrelax.out
-out:file:silent_struct_type binary -abinitio:relax -nstruct 200
-ex1 -ex2 -extrachi_cutoff 0     [ABRELAX]

relax.<exe> -s idealize_2jof.pdb -out:file:silent
2jof_nativerelax.out -out:file:silent_struct_type binary -
database <rosetta_database>  -frag3  aat000_03_05.200_v1_3.txt  -
frag9 aat000_09_05.200_v1_3.txt -native 2jof.pdb -nstruct 200  -
ex1 -ex2 -extrachi_cutoff 0   [NATIVE RELAX]
```

A total of 20,000 models were generated for both *de novo* and native optimization runs. Much larger runs (up to 1,000,000 models; unpub. data, D. Baker & RD) did not give significantly lower energies. Both of these standard command lines are also executable with Rosetta release 3.2.

Independent modeling runs were carried out with a novel StepWise Assembly (SWA) method. A full benchmark of this method is under prepration. Briefly, the method recursively builds each subfragment [*i,j*] of the target sequence onto clustered conformational ensembles (with up to 1000 members, clustered at 0.25 Å) derived from subfragments [*i,j-1*] and [*i+1,j*]. Each "step" involves exhaustively sampling φ/ψ in 20° increments, repacking side-chains, and minimizing. An example command-line for the step building subfragment [3,5] from results for subfragment [4,5]:

```
stepwise_protein_test.<exe> -database <path to rosetta_database>
-rebuild -out:file:silent_struct_type binary  -fasta 2jof.fasta -
n_sample 18 -nstruct 100 -cluster:radius    0.100 -
extrachi_cutoff 0 -ex1 -ex2 -score:weights score12.wts -
pack_weights pack_no_hb_env_dep.wts -add_peptide_plane -native
2jof.pdb -mute all  -silent1 region_4_5_sample.cluster.out -tags1
S_0 -input_res1  4 5 -sample_res   3 4 -out:file:silent
```

```
REGION_3_5/START_FROM_REGION_4_5_DENOVO_S_0/region_3_5_sample.out
```
*[SWA]*

A complete directed acyclic graph (DAG) of the rebuild and clustering steps, along with associated commands in Condor format, was automatically generated by a master Python script. The script and a resulting example DAG are provided via Rosetta protocol capture (see below). The DAG was computed via DAGMAN with the Condor computing platform or with in-house Python scripts (also provided by protocol capture) on the LSF queuing platform on 200 to 400 cores on Stanford's BioX$^2$ resource. Optimized native conformations were also estimated with the StepWise Assembly method. To ensure a fair comparison, the entire calculation was repeated, but using Rosetta atom-pair constraints (with the Rosetta smoothed step function "fade") to keep models with inter-residue Cα-Cα distances within ±1 Å and the tryptophan rotamer in the native conformation. Explicitly, an example command line is:

```
stepwise_protein_test.<exe> -database <path to rosetta_database>
-rebuild -out:file:silent_struct_type binary  -fasta 2jof.fasta -
n_sample 18 -nstruct 100 -cluster:radius     0.100 -
extrachi_cutoff 0 -ex1 -ex2 -score:weights score12.wts -
pack_weights pack_no_hb_env_dep.wts -add_peptide_plane -native
2jof.pdb -mute all  -silent1 region_4_5_sample.cluster.out -tags1
S_0 -input_res1  4 5 -sample_res  3 4 -out:file:silent
REGION_3_5/START_FROM_REGION_4_5_DENOVO_S_0/region_3_5_sample.out
-cst_file 2jof_native_CA_CA_trp.cst   [SWA NATIVE]
```

and the constraint file `2jof_native_CA_CA_trp.cst` is provided by Rosetta protocol capture.

*(B) α-conotoxin GI*

The modeled sequence was: **ECCNPACGRHYSC**. The methods for *de novo* modeling α-conotoxin were essentially the same as for Trp cage. However, the following command lines need to be run from the Das lab branch, which disables complications in disulfide input/output and scoring in the Rosetta release 3.2.

```
AbinitioRelax.<exe> -database <path to rosetta_database> -fasta
1not_.fasta -frag3  aa1not_03_05.200_v1_3  -frag9
aa1not_09_05.200_v1_3  -out:file:silent
1not_abrelax_CST_increase_cycles_no_hb_env_dep.out -
out:file:silent_struct_type binary  -nstruct 400  -cst_file
1not_native_disulf_CEN.cst  -abinitio:relax  -cst_fa_file
1not_native_disulf.cst -native 1not.pdb -increase_cycles 10 -
score:weights score12_no_hb_env_dep.wts  -ex1 -ex2 -
extrachi_cutoff 0      [ABRELAX]
```

and

```
relax.<exe> -database <path to rosetta_database> -s
idealize_1not.pdb -fasta 1not_.fasta -frag3
aa1not_03_05.200_v1_3  -frag9 aa1not_09_05.200_v1_3  -
out:file:silent 1not_native_relax.out -
out:file:silent_struct_type binary  -nstruct 200    -
abinitio:relax  -cst_fa_file 1not_native_disulf.cst -native
1not.pdb -increase_cycles 10 -score:weights score12.wts  -ex1 -
ex2 -extrachi_cutoff 0      [NATIVE RELAX]
```

The constraint file (1not_native_disulf.cst; see protocol capture) enforces near-native disulfide bond lengths and angles between residue pairs (2,7) and (3,13); Rosetta atom-pair constraints are defined, penalizing Sγ-Sγ distances outside 1.5-2.5 Å and inter-residue Sγ-Cβ distances outside 2.5–3.5 Å. For both command-lines, 20,000 models were generated.

The StepWise Assembly command lines were entirely analogous to the ones used for Trp Cage. Explicitly, examples of building subfragment (3,5) from subfragment (4,5) are:

```
stepwise_protein_test.<exe> -database <path to rosetta_database>
-rebuild -out:file:silent_struct_type binary  -fasta 1not.fasta -
n_sample 18 -nstruct 100 -cluster:radius    0.100 -
extrachi_cutoff 0 -ex1 -ex2 -score:weights score12.wts -
pack_weights pack_no_hb_env_dep.wts -add_peptide_plane -cst_file
1not_native_disulf.cst -native 1not.pdb -mute all   -silent1
region_4_5_sample.cluster.out -tags1 S_0 -input_res1   4 5 -
sample_res  3 4 -out:file:silent
REGION_3_5/START_FROM_REGION_4_5_DENOVO_S_0/region_3_5_sample.out
[SWA]
```

```
stepwise_protein_test.<exe> -database <path to rosetta_database>
-rebuild -out:file:silent_struct_type binary  -fasta 1not.fasta -
n_sample 18 -nstruct 100 -cluster:radius    0.100 -
extrachi_cutoff 0 -ex1 -ex2 -score:weights score12.wts -
pack_weights pack_no_hb_env_dep.wts -add_peptide_plane -cst_file
1not_native_disulf_CA_CA.cst -native 1not.pdb -mute all  -silent1
region_4_5_sample.cluster.out -tags1 S_0 -input_res1   4 5 -
sample_res  3 4 -out:file:silent
REGION_3_5/START_FROM_REGION_4_5_DENOVO_S_0/region_3_5_sample.out
```
*[SWA NATIVE]*

*(C) Chymotrypsin inhibitor loop*

The chymotrypsin inhibitor sequence was the 62-residue truncated sequence from barley seeds: TEWPELVGKSVEEAKKVILQDKPEAQIIVLP**VGTIVTMEYRI**DRVRLFVDKLDNIAEVPRVG  (the remodeled loop, residues 35-45 in numbering of PDB 2CI2, is in boldface). The Rosetta command-line made use of a recent loop modeling that leverages kinematic loop closure [35], and is available through Rosetta release 3.2:

```
loopmodel.<exe> -database <path to rosetta_database> -
loops:remodel perturb_kic -loops:refine refine_kic -
loops:input_pdb 2ci2_min.pdb -in:file:native 2ci2.pdb -
loops:loop_file 2ci2_35_45.loop -loops:max_kic_build_attempts
10000 -in:file:fullatom -out:file:fullatom -out:prefix 2ci2 -
out:pdb -ex1 -ex2 —extrachi_cutoff 0 -out:nstruct 200 -
out:file:silent_struct_type binary  -out:file:silent
2ci2_kic_loop35_45.out    [KIC]
```

10,000 KIC models were generated. Output files were rescored to generate RMSDs over just the rebuilt loops, using the command line:

```
score.<exe>  -database <path to rosetta_database> -in:file:silent
2ci2_kic_loop35_45.out -native 2ci2.pdb -out:file:scorefile
2ci2_kic_loop35_45.recalculate_rmsd.sc  -
in:file:silent_struct_type binary -in:file:fullatom -
native_exclude_res 1 2 3 4 5 6 7 8 9 10 11 12 13 14 15 16 17 18
19 20 21 22 23 24 25 26 27 28 29 30 31 43 44 45 46 47 48 49 50 51
52 53 54 55 56 57 58 59 60 61 62
```

The optimized native conformation (2ci2_min.pdb) was generated by packing and minimizing side-chains, as described in [35]. The best resulting energy was 6-7 score units greater than near-native models (< 0.7 Å Cα RMSD) achieved in de novo KIC modeling; the latter energies are given in Fig. 2 and Table 1.

StepWise Assembly methods were also applied to this case, but could not be directly compared to the KIC results because of difference in which degrees of freedom were optimized (the KIC protocol samples N-Cα-C bond angles, for example, whereas the StepWise Assembly code keeps all bond angles fixed).

*(D) UUCG tetraloop (RNA)*

The eight-nucleotide modeled RNA sequence, derived from residues 31-38 of a ribosomal fragment (PDB: 1F7Y), was: **gcuucggc**. (Lower-case letters refer to nucleic acids in Rosetta.)

Fragment Assembly of RNA with Full-Atom Refinement [19] applied the following command line, available in Rosetta release 3.2:

```
rna_denovo.<exe> -random_delay 20 -database  <path to
rosetta_database> -fasta gcuucggc.fasta -nstruct 200 -
out::file::silent gcuucggc.out -minimize_rna -cycles 5000 -mute
all -native gcuucggc_RNA.pdb
```
*[FARFAR]*

Optimized native conformations used a similar command line but drew fragments only from the crystallographic model that was the source of the puzzle:

```
rna_denovo.<exe> -random_delay 20 -database  <path to
rosetta_database>-fasta gcuucggc.fasta -nstruct 200 -
out::file::silent gcuucggc_NATIVE.out -minimize_rna -cycles 5000
-mute all -native gcuucggc_RNA.pdb  -vall_torsions
1f7y_native.torsions
```
*[FARFAR NATIVE]*

In both cases, 20,000 FARFAR models were generated. The native torsion file was generated by:

```
rna_database.<exe>  -database <path to rosetta_database> -s
1f7y_RNA.pdb  -vall_torsions -o 1f7y_native.torsions
```

For RNA modeling cases, StepWise Assembly provides a more efficient sampling method (a full manuscript is in preparation; P. Sripakdeevong & RD, unpub. results). Analogous to protein cases (A) and (B), sub-fragments [*i,j*] of the target sequence are modeled from clustered conformational ensembles for subfragments [*i,j-1*] and [*i+1,j*] in a recursive manner. Single-residues are enumeratively sampled (at, χ, δ, ε, ζ, α, β, and γ) in 20° increments, repacking 2´-OH groups, and minimizing. Here, an ideal Watson-Crick stem was assumed for residues 1-2 and 7-8, and the UUCG loop 3–6 was rebuilt from both ends and connected by CCD loop closure. An example command-lines for the basic rebuild step building residue 3 onto the starting stem in either *de novo* and native-optimization runs are:

```
rna_swa_test.<exe> -algorithm rna_resample_test -database <path
to rosetta_database> -fasta gcuucggc.fasta -output_virtual  -
cluster:radius   0.100  -num_pose_kept 100  -score:weights
rna_hires_2008.wts -native motif2_1f7y_RNA.pdb -
rna_torsion_potential rd2008 -s1 gcgc.pdb  -input_res1 1 2 7 8 -
out:file:silent
REGION_0_1/START_FROM_REGION_0_0/region_0_1_sample.out  -
sample_res 3    [SWA]

rna_swa_test.<exe> -algorithm rna_resample_test -database <path
to rosetta_database> -fasta gcuucggc.fasta -output_virtual  -
cluster:radius   0.100  -num_pose_kept 100  -score:weights
rna_hires_2008.wts -native motif2_1f7y_RNA.pdb  -cst_file
uucg_polar_fade.cst -sampler_native_rmsd_screen -
sampler_native_rmsd_screen_cutoff    1.500    -
rna_torsion_potential rd2008 -s1 gcgc.pdb  -input_res1 1 2 7 8 -
out:file:silent
REGION_0_1/START_FROM_REGION_0_0/region_0_1_sample.out  -
sample_res 3    [SWA NATIVE]
```

In the latter command-line, a constraint file penalizes conformations in which contacting (within 4 Å) polar heavy atoms are placed beyond 1 Å from their native distances; the file is provided in the protocol capture (see next).

**Protocol capture**

All files, including fragments, sequence files (.fasta), native conformations (.pdb), as well as example logs are being provided via "protocol capture" in the Rosetta Subversion repository:

https://svn.rosettacommons.org/source/trunk/RosettaCon2010/protocol_capture/rhiju_four_small_puzzles

The directory will be gladly provided to readers without access to the repository upon request.

## Acknowledgments

I am grateful to all contributors to the Rosetta codebase for sharing their work and to D. Baker, K. Beauchamp, P. Sripakdeevong, and M. Tyka for illuminating discussions.

**Table 1. Comparison of best scoring Rosetta models with optimized experimental models**

|  |  | Lowest energy *de novo* model | | Lowest energy optimized native model | | |
|---|---|---|---|---|---|---|
| Puzzle | Length | energy[a] | rmsd[b] | energy[a] | rmsd[b] | energy gap[c] |
| A. Trp cage | 20 | −40.4 | 2.14 | −38.1 | 0.66 | −2.27 |
| B. α-conotoxin GI | 13 | 8.3 | 2.83 | 10.5 | 0.35 | −2.19 |
| C. chym. inhib. loop | 11 | −102.0 | 1.77 | −98.1 | 0.69 | −3.99 |
| D. UUCG RNA | 4[d] | −63.0 | 4.26 | −53.2 | 0.68 | −9.86 |

[a] Rosetta all-atom energy ("score12") for protein cases A-C [49], and Rosetta FARFAR energy for RNA case D [19]. A Rosetta score unit is approximately 0.5–1 kcal/mol [50].
[b] Cα RMSD for proteins; all-atom RMSD for RNA.
[c] Energy of *de novo* model minus energy of optimized native. A negative sign (observed in all cases) signifies an energy function error.
[d] The entire RNA construct is 8 residues, but only 4 residues are built de novo.

**Figure 1.** Small "puzzles" for high resolution Rosetta tests. (A) Trp cage, (B) α-conotoxin GI, (C) Reactive loop of chymotrypsin inhibitor from barley, (D) the UUCG tetraloop (RNA). Each panel shows experimental structures side-by-side with lowest energy Rosetta *de novo* model discovered in extensive runs (see Fig. 2 and Table 1).

### A. Trp cage

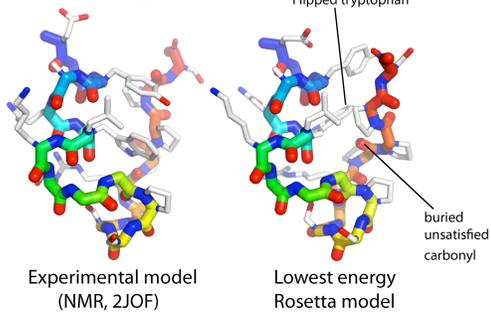

Experimental model (NMR, 2JOF) — Lowest energy Rosetta model (Flipped tryptophan; buried unsatisfied carbonyl)

### B. α-conotoxin GI

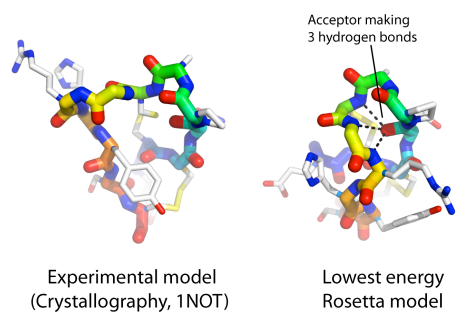

Experimental model (Crystallography, 1NOT) — Lowest energy Rosetta model (Acceptor making 3 hydrogen bonds)

### C. Chymotrypsin inhibitor loop

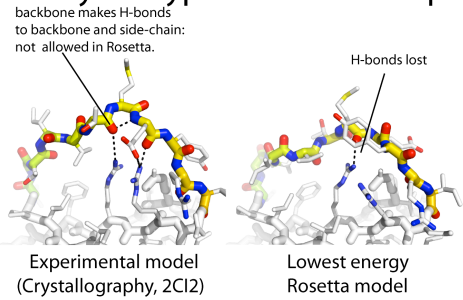

backbone makes H-bonds to backbone and side-chain: not allowed in Rosetta.

Experimental model (Crystallography, 2CI2) — Lowest energy Rosetta model (H-bonds lost)

### D. RNA tetraloop (UUCG)

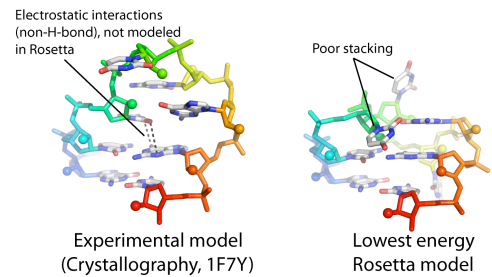

Electrostatic interactions (non-H-bond), not modeled in Rosetta

Experimental model (Crystallography, 1F7Y) — Lowest energy Rosetta model (Poor stacking)

**Figure 2.** All-atom energy vs. RMSD plots for *de novo* modeling of the four puzzles and for optimizing experimental ("native") conformations. Panels correspond to exactly to panels in Fig. 1. In protein cases (A)-(C), the default Rosetta all-atom energy function for *de novo* protein modeling (score12) is plotted against Cα RMSD. In the RNA case (D), the FARFAR energy function (which contains torsional terms for RNA, an orientation-dependent solvation function, and a carbon-hydrogen-bond model [19]) is plotted against all-heavy-atom RMSD. The conformational sampling algorithms (ABRELAX, SWA, etc.) used in the runs are denoted in the figure and described in detail in Methods.

### A. Trp cage
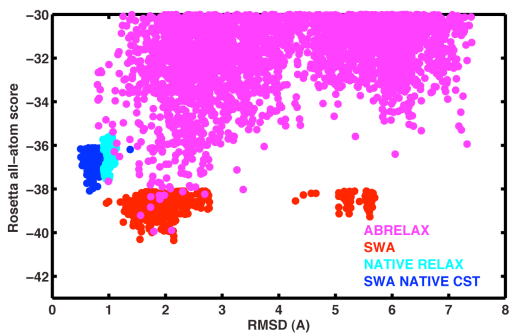

### B. α-conotoxin GI
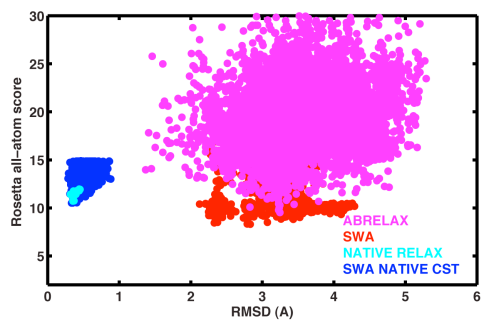

### C. Chymotrypsin inhibitor loop
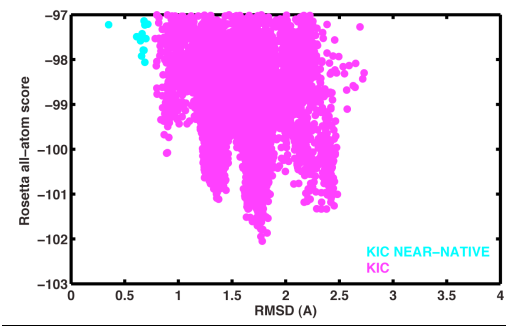

### D. RNA tetraloop (UUCG)
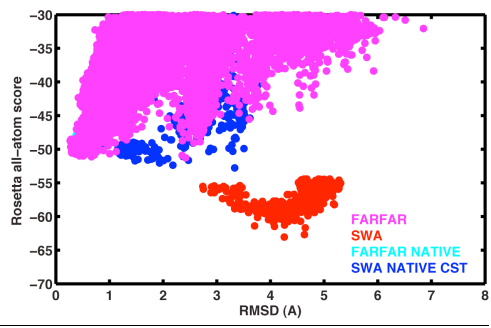